\documentclass[preprint,12pt]{elsarticle}

\usepackage{amssymb}
\usepackage{amsthm}
\usepackage{amsmath}
\usepackage{bm}
\usepackage{physics}
\usepackage{mathtools}
\usepackage{placeins}
\usepackage{subfig}
\usepackage{adjustbox}
\usepackage{multirow}
\usepackage{diagbox}
\usepackage{xcolor}
\usepackage{dsfont}
\theoremstyle{definition}
\newtheorem{definition}{Definition}[section]

\theoremstyle{remark}

\begin{document}

\begin{frontmatter}



\title{A Transferable Physics-Informed Framework for Battery Degradation Diagnosis, Knee-Onset Detection and Knee Prediction}


\author[label1,label2]{Huang Zhang\corref{cor1}}
\ead{huangz@chalmers.se}
\author[label1]{Xixi Liu}
\ead{xixil@chalmers.se}
\author[label2]{Faisal Altaf}
\ead{faisal.altaf@volvo.com}
\author[label1]{Torsten Wik}
\ead{torsten.wik@chalmers.se}
\cortext[cor1]{Corresponding author.}
\affiliation[label1]{organization={Department of Electrical Engineering},%Department and Organization
            addressline={Chalmers University of Technology}, 
            city={Gothenburg},
            postcode={41296}, 
            country={Sweden}}
            
\affiliation[label2]{organization={Department of Electromobility},%Department and Organization
            addressline={Volvo Group Trucks Technology}, 
            city={Gothenburg},
            postcode={40508}, 
            country={Sweden}}

\begin{abstract}
% Move 1 - Background/Introduction/Situation
The techno-economic and safety concerns of battery capacity knee occurrence call for developing online knee detection and prediction methods as an advanced battery management system (BMS) function. 
% Move 2 - Present research purpose
% Move 3 - Methods/Materials/Subjects/Procedures
To address this, a transferable physics-informed framework that consists of a histogram-based feature engineering method, a hybrid physics-informed model, and a fine-tuning strategy, is proposed for online battery degradation diagnosis and knee-onset detection. The hybrid model is first developed and evaluated using a scenario-aware pipeline in protocol cycling scenarios and then fine-tuned to create local models deployed in a dynamic cycling scenario.
% Move 4 - Results/Findings
A 2D histogram-based 17-feature set is found to be the best choice in both source and target scenarios. The fine-tuning strategy is proven to be effective in improving battery degradation mode estimation and degradation phase detection performance in the target scenario. Again, a strong linear correlation was found between the identified knee-onset and knee points.
% Move 5 - Discussion/Conclusion/Implications/Recommendations (no future work)
As a result, advanced BMS functions, such as online degradation diagnosis and prognosis, online knee-onset detection and knee prediction, aging-aware battery classification, and second-life repurposing, can be enabled through a battery performance digital twin in the cloud.
\end{abstract}

\begin{graphicalabstract}
\includegraphics[width=\textwidth]{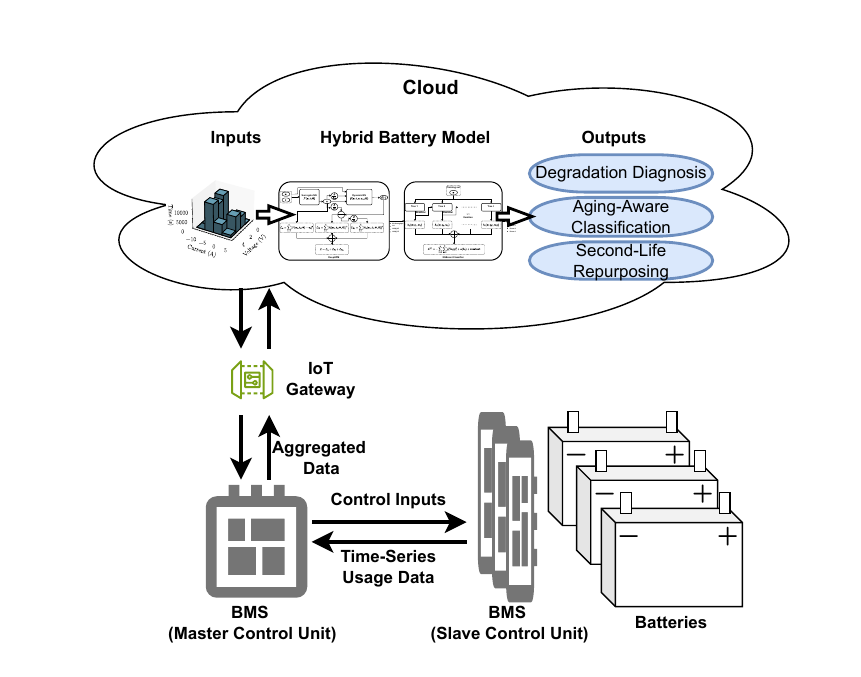}
\end{graphicalabstract}

\begin{highlights}
\item 2D histogram features to enable online battery diagnosis and knee-onset detection.
\item A hybrid model for battery degradation mode estimation and phase detection.
\item A fine-tuning strategy to create local models for online deployment.
\item The method enables advanced battery management system functions in the cloud.
\end{highlights}


\end{frontmatter}


\section{Introduction}\label{sec_introduction}
% Step 1.1 - [Motivation] Establish a territory, showing that the general research problem (i.e., battery capacity knee-onset detection and knee prediction) is meaningful but challenging due to the fact X.
Lithium-ion batteries in electric vehicles and stationary energy storage systems are critical for the cost-effective decarbonization of the transportation and power sectors~\cite{parlikar2023high}~\cite{woody2023decarbonization}. One of the most challenging requirements of these applications is a long battery lifetime to achieve the economic return on investment~\cite{hu2020battery}. However, batteries are designed and used for a wide range of applications, and exhibit path-dependent degradation (i.e., the rate and extent of battery degradation depend not only on operating conditions but also on the specific sequence in the usage history), which leads to considerably dispersed battery lifetime~\cite{gering2011investigation}~\cite{baure2020durability}~\cite{raj2020investigation}. This is caused by complex interactions of various mechanical and chemical degradation mechanisms, many of which are influenced by operating conditions~\cite{reniers2019review}. 
In experimental aging tests of commercial batteries, it is commonly observed that batteries exhibit abrupt capacity fade (also called knee occurrence) which severely limits battery performance and lifetime~\cite{yang2017modeling}~\cite{keil2019linear}~\cite{keil2020electrochemical}~\cite{fang2021capacity}. Moreover, severe safety issues, such as thermal runaway, may arise if batteries are reused after knee occurrence~\cite{gu2024challenges}. Therefore, avoiding or at least delaying knee occurrence is essential to ensure a long battery lifetime.

% Step 1.2 - [Background] Establish a territory, review existing methods that have been used to address the research problem (i.e., battery degradation mode estimation and phase detection); battery applications, problem formulation, model architecture, and shortcomings.
As a key step to avoid or delay knee occurrence throughout a battery's service life, an online capacity knee detection and prediction method, with a possibility of real-time degradation diagnosis and prognosis to unravel why a knee occurs, is sought. As a result, a number of recent research efforts have been made to develop such a method. The data-driven methods focusing on knee detection and prediction aspects can be generally divided into two categories, i.e., intersection-based methods~\cite{diao2019algorithm}~\cite{fermin2020identification}~\cite{greenbank2021automated}, and learning-based methods ~\cite{zhang2019accelerated}~\cite{sohn2022two}~\cite{haris2022degradation}~\cite{you2023nonlinear}~\cite{costa2024icformer}. 
Specifically in the former category, the slope-changing ratio method~\cite{diao2019algorithm}, the Bacon--Watts method~\cite{fermin2020identification}, and the bisector method~\cite{greenbank2021automated} have been proposed, which are based on finding the intersection between a straight line approximating the early fade and a second line approximating the fade after knee occurrence. 
In the latter category, the quantile regression method~\cite{zhang2019accelerated}, convolutional neural networks~\cite{sohn2022two}~\cite{haris2022degradation}, long short-term memory~\cite{you2023nonlinear}, and a transformer-based model~\cite{costa2024icformer} have been proposed, which are based on learning machine learning models with specifically extracted input features from time-series battery data. 
% Step 2.1 - Establish the niche by indicating gaps
In summary, the performance of intersection-based methods is greatly affected by the shape of capacity fade curves which can be linear, sublinear, superlinear or a combination of the three~\cite{attia2022knees}. Moreover, they are not applicable for online detection as they need more or less complete capacity fade curves. In contrast, learning-based methods can be used for online detection and prediction, even with battery degradation diagnosis to some degree~\cite{costa2024icformer}, but they require large amounts of labeled data for model training purposes and are prone to failure when generalizing to usage scenarios unseen at the training stage.

Another important aspect of online capacity knee detection and prediction is non-invasive battery degradation diagnosis, whose methods can be divided into model-based methods~\cite{xiong2018electrochemical}~\cite{kim2021effective}~\cite{fan2023nondestructive}~\cite{teliz2022identification}~\cite{barzacchi2022enabling}~\cite{dubarry2012synthesize}, and curve-based methods~\cite{zhang2020identifying}~\cite{severson2019data}~\cite{birkl2017degradation}~\cite{dubarry2006incremental}~\cite{bloom2005differential}. Specifically in the former category, electrochemical models derived from first principles using porous electrode theory (e.g., the pseudo-two-dimensional model~\cite{xiong2018electrochemical}~\cite{kim2021effective} and the single particle model (SPM)~\cite{fan2023nondestructive}), and equivalent circuit models~\cite{teliz2022identification}~\cite{barzacchi2022enabling} have been proposed. The model parameters that are highly correlated with underlying degradation mechanisms or modes are then identified and tracked for battery degradation diagnosis. Another example is a mechanistic model proposed by Dubarry et al.~\cite{dubarry2012synthesize} that can simulate various "what-if" scenarios of battery degradation modes (e.g., loss of lithium inventory and loss of active material at both electrodes) and enable non-invasive battery degradation diagnosis via incremental capacity (IC) and differential voltage (DV) curves.
In the latter category, curve-based methods that utilize measurements from cell characterization tests, such as electrochemical impedance spectroscopy (EIS)~\cite{zhang2020identifying}, discharge voltage curves~\cite{severson2019data}, pseudo open circuit voltage (OCV)~\cite{birkl2017degradation}, incremental capacity analysis (ICA)~\cite{dubarry2006incremental} and differential voltage analysis (DVA)~\cite{bloom2005differential}, provide alternative solutions to non-invasive battery degradation diagnosis.
% Step 2.2 - Establish the niche by indicating gaps
In summary, model-based methods can simulate path-dependent battery degradation under a range of operating conditions, which can be used for online battery degradation diagnosis with a trade-off between physical accuracy and model complexity. However, there are still many degradation mechanisms that remain poorly understood, and existing physics-based models suffer from poor identifiability, which limits their applications for online degradation diagnosis~\cite{andersson2022parametrization}. In contrast, curve-based methods require either EIS measurements over a frequency range at an electrochemical equilibrium point or pseudo-OCV measurements at a low rate (i.e., C/25 or lower~\cite{barai2019comparison}), which are challenging to acquire in an onboard BMS. 

% Step 3 - Occupy the niche by bridging the gap with this work, i.e., outline the purpose of this work; announce principal findings; state the value of this work.
\textbf{Contribution:} The goal of this work is to fill the gaps indicated above by proposing a transferable physics-informed framework for online battery degradation diagnosis, knee-onset detection, and knee prediction. Specifically, the model takes histograms aggregated from time-series voltage and current data, which are easy to acquire in an onboard BMS. In addition, its performance can be generalized to unseen battery usage scenarios at a satisfactory level using a small amount of labeled data in a target scenario.
Our key results and contributions are as follows:
\begin{itemize}
\item As a result of curvature-based knee and knee-onset identification, the concept of "degradation phases" is introduced, with which batteries can be classified according to which degradation phase they are in. Specifically, Phase 1 is defined as the period from the beginning of life to the knee-onset point in which batteries may continue their usage in first-life applications; Phase 2 is defined as the period from the knee-onset point to the knee point in which batteries may either continue their usage in first-life applications or be repurposed to second-life applications in which the knee occurrence may be avoided; Phase 3 is defined as the period beyond the knee point in which batteries may potentially be repurposed to very mild second-life applications in which the knee occurrence can be stopped, or be recycled.
\item A transferable physics-informed framework is proposed, which consists of a histogram-based feature engineering method, a hybrid physics-informed model, and a fine-tuning strategy. The 2D histogram-based 17-feature set retains the correlation between current and voltage with its predictive power generalizing across different usage scenarios. The proposed hybrid physics-informed model is the first application of a deep hidden physics model (DeepHPM) for battery degradation mode estimation and an XGBoost model for battery degradation phase detection.
\item The hybrid model using a 2D histogram-based 17-feature set is found to be the best choice for estimating battery degradation modes in both source and target scenarios. The fine-tuning strategy was proven to be effective in improving battery degradation mode estimation and phase detection performance in the target scenario. With degradation phases detected with high accuracy, online prediction of battery capacity knee points can also be achieved by leveraging the strong linear correlations identified between knee-onset and knee points.
\end{itemize}

\section{Methods}
% Summarize each subsection
\subsection{Related definitions}
\begin{definition}[Degradation mechanisms~\cite{birkl2017degradation}~\cite{vetter2005ageing}]\label{def2_1}
    A degradation mechanism is a mechanical or chemical mechanism that degrades the different components of a battery, such as the electrodes, the electrolyte, the separator, and the current collectors. In this work, three degradation mechanisms are considered, namely, solid electrolyte interphase (SEI) growth, lithium plating, and particle cracking.
\end{definition}

\begin{definition}[Degradation modes~\cite{birkl2017degradation}~\cite{pastor2016identification}]\label{def2_2}
    A degradation mode is a degradation mechanism that has a unique and measurable effect on, for example, the capacity, impedance, and open circuit voltage (OCV) of lithium-ion cells, caused by one or multiple interacting degradation mechanisms. Three degradation modes are considered here, loss of lithium inventory (LLI), loss of active material at the negative electrode (LAM\_NE), and loss of active material at the positive electrode (LAM\_PE). Notably, another degradation mode, i.e., conductivity loss (CL), describes ohmic resistance increase.
\end{definition}

\begin{definition}[Knee~\cite{IEEE485}]
    The capacity knee is defined as the point when "the capacity slowly declines throughout most of the battery’s life, but begins to decrease rapidly in the latter stages" (see Fig.~\ref{fig1_1}).
\end{definition}

\begin{definition}[Knee-onset~\cite{fermin2020identification}]
    The capacity knee-onset is defined as "the point that marks the beginning of the accelerated degradation rate at which the capacity fade can no longer be approximated as a linear function". The knee-onset point shows the very first signs of the transition to the accelerated degradation rate, which provides much earlier warning than the knee point, where the accelerated degradation is already taking place (see Fig.~\ref{fig1_1}).
\end{definition}

\begin{definition}[Degradation phases~\cite{zhang2024battery}]\label{def2_3}
    A battery degradation process with knee occurrence on the capacity fade curve has three discrete phases $\bm{S} = [s_1, s_2, s_3]$ separated by two boundaries $\bm{B} = [b_1, b_2]$. Here, the first phase ($s_1$) represents the battery degradation process from the beginning of life to the knee-onset point ($b_1$), the second phase ($s_2$) represents the battery degradation process from the knee-onset point ($b_1$) to the knee point ($b_2$), and the third phase ($s_3$) represents the battery degradation process from the knee point ($b_2$) to the end of life.
\end{definition}

\begin{definition}[Path-dependent degradation~\cite{gering2011investigation}]
    The rate and extent of degradation depend not only on aging conditions but also on the specific sequence of aging conditions in the usage history.
\end{definition}

\begin{definition}[Knee pathways~\cite{attia2022knees}]\label{def2_4}
    A knee pathway is a family of battery internal state (associated with degradation mechanisms) trajectories that lead to a knee on the capacity fade curve, such as lithium plating, electrode saturation, resistance growth, electrolyte and additive depletion, percolation-limited connectivity, and mechanical deformation.
\end{definition}

\subsection{Aging-aware battery classification}\label{subsec3_1}
% Motivate aging-aware battery classification
Lithium-ion batteries are designed, manufactured, and used for a wide range of applications, such as portable electronics, electric vehicles (EVs), and stationary energy storage systems. As a result of complex interactions between multiple degradation mechanisms that are influenced by operating conditions (e.g., charge/discharge current, temperature, state-of-charge (SoC) window, etc.), batteries can exhibit capacity fade curves that are linear, sublinear, superlinear or a combination of the three~\cite{attia2022knees}. In particular, superlinear capacity fade infers a knee on the capacity fade curve, which may significantly shorten a battery's lifetime and pose safety risks~\cite{gao2024comprehensive}. Therefore, batteries with knee occurrence are recommended to retire immediately for safety concerns~\cite{martinez2018technical}. As a result, retired EV batteries with knee occurrence will have no second-life value. However, if we can classify EV batteries during their first-life applications, and then repurpose them to well-controlled second-life applications in which the knee occurrence can be delayed or even avoided, then enormous quantities of EV batteries could potentially have substantial second-life value with guaranteed safety.

\begin{figure*}[!ht]
\centering
\subfloat[The capacity fade curve with 3 identified degradation phases (top) and estimated degradation modes (bottom).]{\includegraphics[width=2.5in]{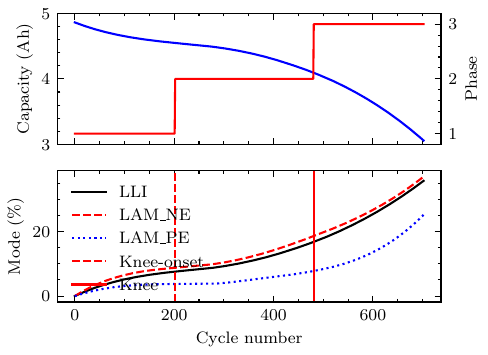}%
\label{fig1_1}}
\hfil
\subfloat[The capacity fade curve with 2 identified degradation phases (top) and estimated degradation modes (bottom).]{\includegraphics[width=2.5in]{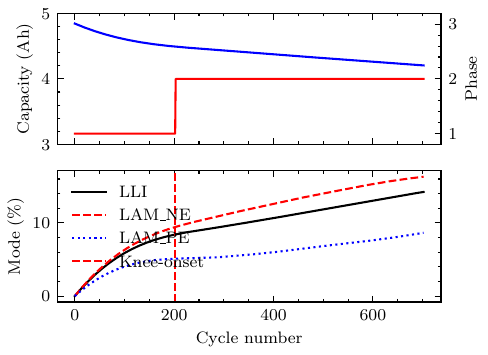}%
\label{fig1_2}}
\caption{The identified degradation phases and estimated degradation modes of cell E4B (left) and E4E (right) in the ICL dataset.}
\label{fig1}
\end{figure*}

% Summarize our previously proposed curvature-based identification method
Our previous work proposed a curvature-based identification method to identify knee-onset and knee points on the battery capacity fade curve~\cite{zhang2024battery}. Identified knee-onset and knee points divide the battery degradation process into three phases (see Definition~\ref{def2_3}), and then batteries can be classified based on which degradation phase they are in:
\begin{itemize}
\item[] 
\textbf{Phase 1}: From the beginning of life to the knee-onset point. Batteries in this phase may continue their usage in first-life applications.
\item[] 
\textbf{Phase 2}: From the knee-onset point to the knee point. Batteries in this phase may either continue their usage in first-life applications or be repurposed to very mild second-life applications in which the knee occurrence may be avoided or at least significantly delayed.
\item[] 
\textbf{Phase 3}: From the knee point to the end of life. Batteries in this phase may either be repurposed to second-life applications in which the knee occurrence can be stopped, or be recycled.
\end{itemize}
% Demonstrate how it can facilitate the classification of EV batteries
To demonstrate how the curvature-based identification method facilitates the classification of batteries, we use two sample cells, one with knee occurrence and one without knee occurrence, from the Imperial College London (ICL) dataset~\cite{kirkaldy2024lithium}. As illustrated in Fig.~\ref{fig1}, it can be seen that at the end of the experiments cell E4B is in Phase 3 while cell E4E is in Phase 2. Since the knee on the capacity fade curve can be caused by different knee pathways (see Definition~\ref{def2_4}), it is natural to correlate identified degradation phases with estimated degradation modes (see Definition~\ref{def2_2}). The degradation of cell E4B begins with a square root dependence on time until it reaches the knee-onset point (201 cycles) at which its degradation process transits from Phase 1 to Phase 2, and then continues growing until the knee point (478 cycles) at which the degradation process transits from Phase 2 to Phase 3. In contrast, the degradation of cell E4E also begins with a square root dependence on time until the knee-onset point (201 cycles) at which its degradation process transits from Phase 1 to Phase 2. However, there is no knee occurrence before the end of the experiment, and thus, Phase 3 is never reached.

\subsection{Histogram-based feature engineering}
To achieve online battery degradation mode estimation and phase detection, the objective of feature engineering is to reduce the dimensionality of time-series usage data and generate aging-relevant features. A simple histogram-based feature engineering method was found to be able to extract features from time-series usage data, whose excellent predictive power can be generalized across different battery usage scenarios~\cite{greenbank2021automated}~\cite{zhang2024scenario}.

\begin{figure*}[!ht]
\centering
\subfloat[Voltage histogram.]{\includegraphics[width=2.5in]{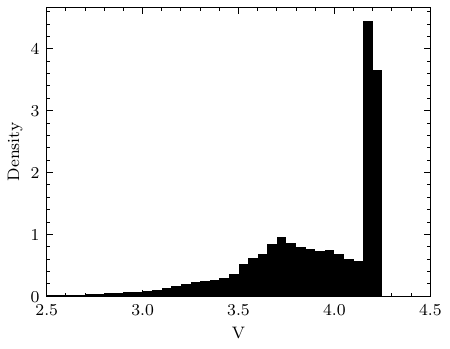}%
\label{fig2_1}}
\hfil
\subfloat[Voltage cumulative histogram.]{\includegraphics[width=2.5in]{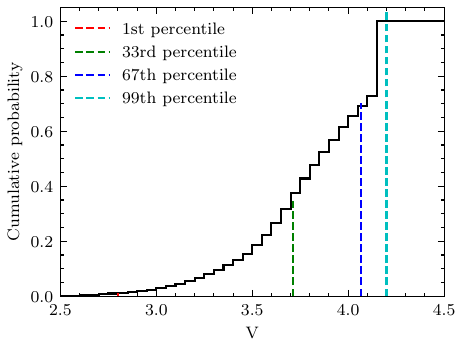}%
\label{fig2_2}}
\caption{The histogram (left) and cumulative histogram (right) generated using the voltage data in Experiment 5 of ICL dataset.}
\label{fig2}
\end{figure*}

\begin{figure*}[!ht]
\centering
\subfloat[Current histogram.]{\includegraphics[width=2.5in]{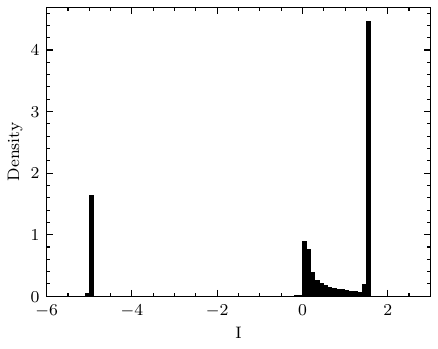}%
\label{fig3_1}}
\hfil
\subfloat[Current cumulative histogram.]{\includegraphics[width=2.5in]{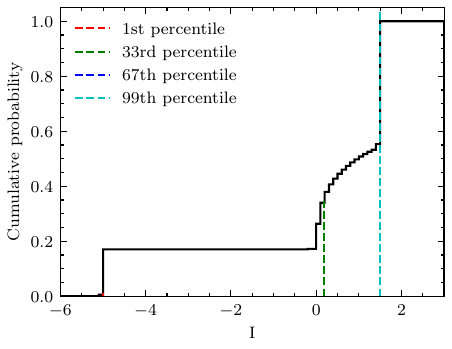}%
\label{fig3_2}}
\caption{The histogram (left) and cumulative histogram (right) generated using the current data in Experiment 5 of ICL dataset.}
\label{fig3}
\end{figure*}

This histogram-based feature engineering method consists of two steps, i.e., first determining the variable bounds; and then extracting features from time-series usage data. We assume that voltage and current measurements are available for each cell in a battery pack, we will therefore focus on features extracted from the time series of these two variables in this work. Specifically, we first determine the variable bounds, for which a histogram and a cumulative histogram are generated from the measured voltage and current data. An example of the histogram and the cumulative one generated from the voltage data can be seen in Fig.~\ref{fig2_1} and~\ref{fig2_2}, respectively. Similarly, an example of the histogram and the cumulative one generated using the current data can be seen in Fig.~\ref{fig3_1} and~\ref{fig3_2}, respectively. Each bar in the histograms represents the time spent within a specific voltage or current range. The 1st, 33rd, 67th, and 99th percentiles of each variable are given in Table~\ref{tab_1}. Based on these bounds, a voltage-based 3-feature set was selected as the optimal set for online capacity estimation by Greenbank and Howey~\cite{greenbank2021automated}. Although battery capacity fade is attributed to the growth of internal degradation modes, this feature set may not be the best feature set for online battery degradation mode estimation.
Moreover, time spent in extreme ranges, for example, ranges outside charge/discharge cut-off voltages, should also be monitored due to its evident degradation effects~\cite{ouyang2022sensitivities}. As listed in Table~\ref{tab_2}, five feature sets are generated for online battery degradation mode estimation.

\begin{table*}[ht]
\caption{Example variable bounds for cells in Experiment 5 of ICL dataset.}
\label{tab_1}
% \begin{adjustbox}{width=\textwidth,center}
\begin{center}
\begin{tabular}{|l|c|c|}
\hline
\textbf{Percentile} & \textbf{Voltage (V)} & \textbf{Current (A)}\\
\hline
1st     & 2.803  & -5 \\
33rd    & 3.711  & 0.181  \\
67th    & 4.068  & 1.501 \\
99th    & 4.200  & 1.505 \\
\hline
\end{tabular}
\end{center}
% \end{adjustbox}
\end{table*}

\begin{table*}[htpb]
\caption{Five feature sets using histogram-based method}
\label{tab_2}
\begin{adjustbox}{width=\textwidth,center}
% \begin{center}
\begin{tabular}{|l|l|}
\hline
\textbf{Feature set} & \textbf{Input feature} \\
\hline
\multirow{3}{*}{1D voltage-based 3-feature set} & \shortstack[l]{Time spent between 1st and 33rd voltage percentiles (i.e., $V_{12}$)} \\ \cline{2-2}
&  \shortstack[l]{Time spent between 33rd and 67th voltage percentiles (i.e., $V_{23}$)}  \\ \cline{2-2}
&  Calendar time (i.e., $t$)  \\
\hline
\multirow{5}{*}{1D voltage-based 5-feature set} & \shortstack[l]{Time spent less than 1st voltage percentile (i.e., $V_{01}$)} \\ \cline{2-2}
& \shortstack[l]{Time spent between 1st and 33rd voltage percentiles (i.e., $V_{12}$)} \\ \cline{2-2}
&  \shortstack[l]{Time spent between 33rd and 67th voltage percentiles (i.e., $V_{23}$)} \\ \cline{2-2}
& \shortstack[l]{Time spent greater than 67th voltage percentile (i.e., $V_{34}$)} \\ \cline{2-2}
&  Calendar time (i.e., $t$) \\
\hline
\multirow{3}{*}{1D current-based 3-feature set} & \shortstack[l]{Time spent between 1st and 33rd current percentiles (i.e., $I_{12}$)} \\ \cline{2-2}
&  \shortstack[l]{Time spent between 33rd and 67th current percentiles (i.e., $I_{23}$)}  \\ \cline{2-2}
&  Calendar time (i.e., $t$)  \\
\hline
\multirow{5}{*}{1D current-based 5-feature set} & \shortstack[l]{Time spent less than 1st current percentile (i.e., $I_{01}$)} \\ \cline{2-2}
& \shortstack[l]{Time spent between 1st and 33rd current percentiles (i.e., $I_{12}$)} \\ \cline{2-2}
&  \shortstack[l]{Time spent between 33rd and 67th current percentiles (i.e., $I_{23}$)} \\ \cline{2-2}
& \shortstack[l]{Time spent greater than 67th current percentile (i.e., $I_{34}$)} \\ \cline{2-2}
&  Calendar time (i.e., $t$) \\
\hline
\multirow{5}{*}{2D current-voltage 17-feature set} & \shortstack[l]{Time spent less than 1st current percentile and in 4 voltage ranges \\ (i.e., $I_{01}V_{01}$, $I_{01}V_{12}$, $I_{01}V_{23}$, $I_{01}V_{34}$)} \\ \cline{2-2}
& \shortstack[l]{Time spent between 1st and 33rd current percentiles and in 4 voltage ranges \\ (i.e., $I_{12}V_{01}$, $I_{12}V_{12}$, $I_{12}V_{23}$, $I_{12}V_{34}$)} \\ \cline{2-2}
&  \shortstack[l]{Time spent between 33rd and 67th current percentiles and in 4 voltage ranges \\(i.e., $I_{23}V_{01}$, $I_{23}V_{12}$, $I_{23}V_{23}$, $I_{23}V_{34}$)} \\ \cline{2-2}
& \shortstack[l]{Time spent greater than 67th current percentile and in 4 voltage ranges \\(i.e., $I_{34}V_{01}$, $I_{34}V_{12}$, $I_{34}V_{23}$, $I_{34}V_{34}$)} \\ \cline{2-2}
&  Calendar time (i.e., $t$) \\
\hline
\end{tabular}
% \end{center}
\end{adjustbox}
\end{table*}

\subsection{Hybrid model architecture design}
% Motivate why degradation modes are modeled for each knee pathway
A knee on the capacity fade curve may be contributed by multiple degradation modes, and each of these modes may be contributed by multiple degradation mechanisms. Although it can be extremely challenging to independently identify each degradation mechanism due to direct and indirect interactions between them, degradation modes are quantifiable using pseudo-OCV measurements and the degradation model proposed by Birkl et al.~\cite{birkl2017degradation}.

% Introduce physics-based modeling approach for degradation modes and its drawbacks
Specifically, SEI growth can contribute to LLI by immobilizing lithium ions; lithium plating can contribute to LLI by forming dead lithium; particle cracking can contribute to LLI by creating new surfaces for SEI growth and lithium plating, and it can also contribute to stress-driven LAM through island formation and binder decomposition. Therefore, modeling LLI requires modeling these three degradation mechanisms and their interactions, while modeling stress-driven LAM requires modeling one degradation mechanism. 
Moreover, depending on the root cause of the capacity knee, the degradation mechanisms contributing to each degradation mode may also differ in each knee pathway (see Definition~\ref{def2_4}). For example, in a cracking-induced knee pathway, the intercalation and deintercalation of lithium ions during cycling can cause alternating mechanical stress in the electrodes, which can lead to particle cracking. New surfaces can be created for SEI growth as the cracks propagate, which accelerates LLI. The accelerated LLI contributes to the accelerated capacity fade, and eventually, a knee on the capacity fade curve. Therefore, modeling LLI in a cracking-induced knee pathway requires modeling two degradation mechanisms (SEI growth and particle cracking) and their interactions. However, the exact degradation mechanisms that contribute to LLI and LAM are not known as a prior for each cell, which makes it almost impossible to model LLI and LAM accurately enough for online degradation mode estimation.

% Motivate the hybrid physics-informed neural network modeling approach (i.e., DeepHPM) for degradation mode estimation.
In this study, to capture key dynamics in each knee pathway, we propose to model each degradation mode as a multivariate function
\begin{equation}\label{eqn_u_solution}
    u_i = f_i(t, \bm{x}), \ i=1,2,3,
\end{equation}
where $t \in \mathbb{R}$ denotes time, $\bm{x} \in \mathbb{R}^m$ denotes an input vector and can be one of five feature sets, excluding the calendar time $t$ (see Table~\ref{tab_2}), and $u_i \in \mathbb{R}$ denotes each of the three degradation modes (see Definition~\ref{def2_2}). The degradation mode growth rate is defined as how quickly a degradation mode grows with respect to time, and can be described using a nonlinear partial differential equation (PDE) in the general form
\begin{equation}\label{eqn_u_dynamics}
    u_t = \pdv{u}{t} = g(t, \bm{x}, u, u_{\bm{x}}),
\end{equation}
where we have omitted the index $i$ to simplify, $g(\cdot)$ is a nonlinear function of time $t$, the input vector $\bm{x}$, solution $u$, and its derivatives with respect to the input vector, for example, $ u_{\bm{x}} = [\pdv{u}{x_1}, \pdv{u}{x_2}, \ldots, \pdv{u}{x_m}]^T$. The function $g(\cdot)$ comprises battery internal degradation dynamics and can represent different forms of degradation, such as linear, sublinear, superlinear, or combination of the three~\cite{attia2022knees}. However, the explicit form of $g(\cdot)$ is unknown and almost impossible to obtain. Inspired by the deep hidden physics model (DeepHPM) proposed by Raissi et al.~\cite{raissi2018deep}, we approximate the function $f(\cdot)$ in Eqn. (\ref{eqn_u_solution}) and the nonlinear function $g(\cdot)$ in Eqn. (\ref{eqn_u_dynamics}) with two neural networks (NNs) and define a DeepHPM $\mathcal{H}$ to model battery degradation mode:
\begin{equation}\label{eqn_DeepHPM_mode}
    \mathcal{H}(t, \bm{x}; \Phi, \Theta) \coloneqq \pdv{\mathcal{F}(t, \bm{x}; \Phi)}{t} - \mathcal{G}(t, \bm{x}, u, u_{\bm{x}}; \Theta), 
\end{equation}
where $\mathcal{F}(\cdot)$ denotes the surrogate NN that approximates the hidden solution of the dynamical models, $\mathcal{G}(\cdot)$ denotes the dynamic NN that approximates the battery degradation dynamics, and $\pdv{\mathcal{F}(t, \bm{x}; \Phi)}{t}$ denotes the partial derivatives of the surrogate NN $\mathcal{F}(\cdot)$ with respect to $t$. Notably, we only consider the first-order partial derivatives in the dynamic NN, to achieve a good trade-off between accuracy and complexity. The parameters $\Phi$ of the surrogate NN $\mathcal{F}(\cdot)$ and $\Theta$ of the dynamic NN $\mathcal{G}(\cdot)$ can be learned by minimizing the sum of squared errors by the loss function
\begin{equation}\label{DeepHPM_loss}
    \mathcal{L}   = \mathcal{L}_u + \mathcal{L}_\mathcal{H} + \mathcal{L}_{\mathcal{H}_t},
\end{equation}
where
\begin{align}
    \mathcal{L}_u = &\sum^{n}_{i=1}[{\mathcal{F}}(t_i, \boldsymbol{x}_i;\Phi) - u_i]^2\\
    \mathcal{L}_\mathcal{H} = &\sum^{n}_{i=1}[{\mathcal{H}}(t_i, \boldsymbol{x}_i; \Phi, \Theta)]^2\\
    \mathcal{L}_{\mathcal{H}_t} = &\sum^{n}_{i=1}[{\mathcal{H}}_t(t_i, \boldsymbol{x}_i; \Phi, \Theta)]^2
\end{align}
where $n$ is the number of training samples and $\mathcal{H}_t = \pdv{\mathcal{H}}{t}$. The data loss term $\mathcal{L}_u$ aims to find the parameters of the surrogate NN $\mathcal{F}(\cdot)$ so that it fits the data, while the PDE loss term $\mathcal{L}_\mathcal{H}$ aims to find the parameters of the dynamic NN $\mathcal{G}(\cdot)$ so that it satisfies the PDE defined by Eqn. (\ref{eqn_u_dynamics}) at the evaluated points $(t_i, \boldsymbol{x}_i)$. Moreover, it has been empirically demonstrated that embedding the gradient of the PDE residual into the loss function can further improve the accuracy of the DeepHPM~\cite{yu2022gradient}. Therefore, the PDE gradient loss term $\mathcal{L}_{\mathcal{H}_t}$ that aims to reduce fluctuations and makes the PDE residual closer to zero is also added here. During the training process, the derivatives of the surrogate NN w.r.t. time $t$ and input $\bm{x}$, and the derivatives of the DeepHPM function w.r.t. time $t$ are evaluated using automatic differentiation~\cite{baydin2018automatic}.

% Motivate the data-driven approach (i.e., XGBoost) for degradation phase detection.
With the availability of battery degradation modes estimated by the DeepHPM, the degradation phase of a battery can be detected. Mathematically, the degradation phase detection can be formulated as a multi-class classification problem to classify each battery into one of the three possible classes (see Subsection~\ref{subsec3_1}) given its estimated degradation modes and calendar time. As an efficient and scalable machine learning system for tree boosting, proposed by Chen and Guestrin~\cite{chen2016xgboost}, XGBoost has demonstrated outstanding performance in battery state of health estimation~\cite{song2020lithium}~\cite{sun2024soh}, and remaining useful life prediction~\cite{jafari2022xgboost}. Thus, it is reasonable to explore its potential for addressing the classification problem here.

In practice, the XGBoost model can be learned by minimizing a general loss function at each boosting iteration, which consists of a training loss term $l(\cdot)$ and a regularization term $\omega(\cdot)$
\begin{equation}\label{XGBoost_loss}
    \mathcal{L}^{(t)} = \sum^{n}_{i=1} l(y_i, \hat{y}^{(t-1)}_i + h_t(\bm{v}_i)) + \omega(h_t) + \mathrm{constant},
\end{equation}
where $n$ is the number of training samples, $y_i$ is the value of the $i$-th sample, $\hat{y}^{t-1}_i$ is the prediction of the $i$-th sample up to the $(t-1)$-th iteration, $h_t(\bm{v}_i)$ is the output of the $t$-th tree for the $i$-th sample. In our case, the XGBoost model is to approximate a function that predicts the class $y \in \{1,2,3\}$ given an input vector $\bm{v} = [u_1, u_2, u_3, t]^T$. Note that $u_1$, $u_2$, and $u_3$ represent three "true" degradation modes governed by three Eqn.~(\ref{eqn_u_solution}), respectively. Thus, the general loss function in Eqn. (\ref{XGBoost_loss}) must be modified to account for multi-class classification as
\begin{equation}
    \mathcal{L}^{(t)} = \sum^{n}_{i=1} \sum^{3}_{k=1} l(y^k_i, \hat{y}^{k,(t-1)}_i + h^k_t(\bm{v}_i)) + \omega(h_t) + \mathrm{constant},
\end{equation}
where the multi-class loss for a sample is typically a categorical cross-entropy loss given by
\begin{equation}
    {l}(y_i, \hat{y}_i) = - \sum^{3}_{k=1}{y^k_i} \log \hat{p}^k_i,
\end{equation}
where $y_i$ is the observed class label of sample $i$, $y^k_i \in \{0,1\}$ indicates whether the observed class of the $i$-th sample is class $k$, $\hat{p}^k_i$ is the predicted probability of the $i$-th sample belonging to class $k$ calculated using the softmax function:
\begin{equation}
    \hat{p}^k_i = \frac{\exp(\hat{y}^{k,t}_i)}{\sum^{3}_{j=1} \exp(\hat{y}^{j,t}_i)}.
\end{equation}
Therefore, the loss function for a 3-class XGBoost classifier can be written as
\begin{equation}\label{XGBoost_classifier_loss}
    \mathcal{L}^{(t)} = -\sum^{n}_{i=1} \sum^{3}_{k=1} y^k_i \log \hat{p}^k_i + \omega(h_t) + \mathrm{constant}.
\end{equation}
% The classifier then outputs the class with the highest probability given a test sample $\bm{v}$,
% \begin{equation}
%     \hat{y} = \argmax_k p_k(\bm{v}).
% \end{equation}
The resulting hybrid physics-informed model is illustrated in Fig.~\ref{fig4},~\ref{appendix_A}.

\subsection{Transfer learning-based battery degradation mode estimation}
The battery degradation process is best understood and most studied in a laboratory environment, in which cycling conditions can be closely controlled and reference performance tests (RPTs) can be periodically conducted to characterize the battery degradation. With high-quality laboratory data, the hybrid physics-informed model (see Fig.~\ref{fig4},~\ref{appendix_A}) can be first developed in specific cycling scenarios (or source scenarios). However, different and even uncontrollable usage scenarios still pose a major challenge to deploying the trained model for online battery degradation mode estimation and phase detection in target scenarios. To address this, transfer learning (TL) is emerging as a promising strategy for transferring existing knowledge from different but related domains to a target domain in the field of advanced battery management~\cite{liu2023transfer}.

% Motivate the model updating for DeepHPM 
In this study, a fine-tuning strategy is proposed to create local models deployed in a target scenario. For the same type of battery, we assume that the parameters in Eqn.~(\ref{eqn_u_dynamics}) remain unchanged across different usage scenarios, while the parameters in Eqn.~(\ref{eqn_u_solution}) vary with battery usage scenarios. Some physical parameters in Eqn. (\ref{eqn_u_dynamics}) may indeed change significantly with battery aging, which is contradictory to this assumption. However, adapting the parameters in both models is computationally demanding and also increases the need of labeled data in a target scenario. Our experience with the DeepHPM also indicates that fine-tuning both the surrogate and dynamical NNs achieves performance comparable to fine-tuning the surrogate NN alone. Thus, the battery degradation modes in the target scenario could be estimated with satisfactory performance by freezing the dynamic NN $\mathcal{G}(\cdot)$ and fine-tuning the surrogate NN $\mathcal{F}(\cdot)$ using only a small amount of labeled data obtained from maintenance in the target scenario. In this way, the pre-trained DeepHPM using a large amount of data in the source scenario can be adapted to a target scenario that differs from the source.

\subsection{Experimental design}
\subsubsection{Dataset description}
\begin{figure}[!ht]
\centerline{\includegraphics[width=0.5\textwidth]{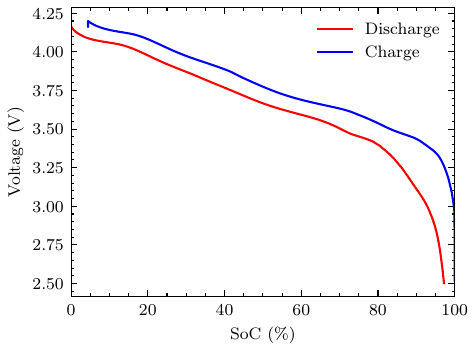}}
\caption{The C/10 discharge-charge curve of a sample cell.}
\label{fig_R1}
\end{figure}

% Description of each battery dataset consists of:
% 1. Institutes info;
% 2. Battery info, i.e., number of cells, cathode chemistry, form factor, manufacturer, model number, nominal capacity, batch date (optional);
% 3. Battery test info, i.e., test purpose (optional), test conditions (e.g., charge/discharge profiles, ambient temperature, SoC window, DoD, etc.);
% 4. End of experiment threshold and lifetime measure (e.g., cycle number, EFC, etc.) in this dataset;
% 5. Available measurements i.e., time-series battery usage in each cycle and battery health per cycle and how they are used in this work
% 6. Rationalize knee occurrence on capacity fade curves in the dataset if there is.
The battery dataset was generated by Imperial College London and the Faraday Institution~\cite{kirkaldy2024lithium}. This high-quality open-source dataset consists of 40 lithium nickel manganese cobalt oxide (NMC 811)/graphite-$\text{SiO}_\text{x}$ cylindrical cells manufactured by LG Chem (model GBM50T2170, 5 Ah nominal capacity with a lower voltage cut-off of 2.5 V and an upper voltage cut-off of 4.2
V). The test aims to characterize battery degradation behaviors (i.e., capacity fade, resistance growth, and degradation mode analysis) under 15 different operating conditions (i.e., ambient temperature, SoC window, and discharge profile) throughout 5 experiments. Specifically, all the cells were charged with a 0.3C constant-current and constant-voltage (CC-CV) charging step, and then discharged with a 1C CC discharging step except for Experiment 4 in which cells were discharged with the World wide harmonized Light vehicle Test Protocol (WLTP) driving profile. Two battery health metrics, i.e., capacity (C/10 discharge, 25$^{\circ}$C) and internal resistance were measured periodically from RPTs. Moreover, three degradation modes, i.e., LLI, LAM\_NE, and LAM\_PE, were also quantified using pseudo-OCV measurements.
Specifically, the pseudo-OCV data for degradation mode analysis was obtained using C/10 discharge-charge cycles at 25$^{\circ}$C between the voltage limits (2.5 V and 4.2 V). The C/10 discharge-charge curve of a sample cell is illustrated in Fig.~\ref{fig_R1}, and the sampling interval of discharge-charge data is 10 seconds. Considering the well-controlled low-rate cycling conditions and the good quality of measured OCV data, we, therefore, took the quantified degradation modes as the "true" degradation modes in our work. To address possible cumulative errors or amplification effects in the model output, we will incorporate uncertainty quantification into our model in future work.

\begin{figure}[!ht]
\centerline{\includegraphics[width=0.5\textwidth]{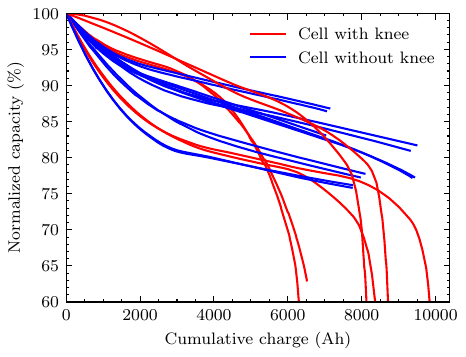}}
\caption{Normalized capacity fade curves of 18 cells in the ICL dataset. Note that capacity (C/10 discharge, 25$^{\circ}$C) from RPTs are used here.}
\label{fig5}
\end{figure}

In this study, 6 cells in Experiment 1 and 6 cells in Experiment 5 are selected for source scenarios as they were aged under two different cycling protocols, and 6 cells in Experiment 4 are selected for the target scenario as they were aged under a dynamic cycling protocol (see Table~\ref{tab_3}). Their capacity fades are illustrated in Fig.~\ref{fig5}. It can be seen that 6 of the cells (E1A, E1B, E1E, E1F, E4B, and E4C) have capacity knee occurrence, which is likely due to lithium plating at the negative electrode as they were cycled at low temperatures~\cite{kirkaldy2024lithium}.
Note that the charging protocols in both source and target scenarios are the same, i.e., 0.3C CC-CV charge, while the discharging protocol in source scenarios (1C galvanostatic discharge) differs from that in the target scenario (WLTP discharge). Therefore, histogram features extracted in source scenarios cannot encode cycling conditions that lead to data leakage.
\begin{table*}[ht]
\caption{Summary of cells}
\label{tab_3}
\begin{adjustbox}{width=\textwidth,center}
% \begin{center}
\begin{tabular}{|l|c|c|c|c|}
\hline
\textbf{Scenario} & \textbf{Charge/Discharge profile} & \textbf{Cell} & \textbf{Ambient temperature} & \textbf{Knee occurrence}\\
\hline
\multirow{6}{*}{\shortstack[c]{Cycling protocol A\\(Source scenario)}} & \multirow{6}{*}{\shortstack[c]{0.3C CC-CV charge/1C galvanostatic discharge\\ 0-30\% SoC window}} & E1A & 10$^{\circ}$C & Yes\\ \cline{3-5}
& & E1B & 10$^{\circ}$C & Yes\\ \cline{3-5}
& & E1E & 25$^{\circ}$C & Yes\\ \cline{3-5}
& & E1F & 25$^{\circ}$C & Yes\\ \cline{3-5}
& & E1K & 40$^{\circ}$C & No\\ \cline{3-5}
& & E1L & 40$^{\circ}$C & No\\
\hline
\multirow{6}{*}{\shortstack[c]{Cycling protocol B\\(Source scenario)}} & \multirow{6}{*}{\shortstack[c]{0.3C CC-CV charge/1C galvanostatic discharge \\ 0-100\% SoC window}} & E5B & 10$^{\circ}$C & No\\ \cline{3-5}
& & E5C & 10$^{\circ}$C & No\\ \cline{3-5}
& & E5D & 25$^{\circ}$C & No\\ \cline{3-5}
& & E5E & 25$^{\circ}$C & No\\ \cline{3-5}
& & E5F & 40$^{\circ}$C & No\\ \cline{3-5}
& & E5G & 40$^{\circ}$C & No\\
\hline
\multirow{6}{*}{\shortstack[c]{Dynamic cycling\\(Target scenario)}} & \multirow{6}{*}{\shortstack[c]{0.3C CC-CV charge/WLTP discharge \\ 0-100\% SoC window}} & E4B & 10$^{\circ}$C & Yes\\ \cline{3-5}
& & E4C & 10$^{\circ}$C & Yes\\ \cline{3-5}
& & E4D & 25$^{\circ}$C & No\\ \cline{3-5}
& & E4E & 25$^{\circ}$C & No\\ \cline{3-5}
& & E4F & 40$^{\circ}$C & No\\ \cline{3-5}
& & E4G & 40$^{\circ}$C & No\\
\hline
\end{tabular}
% \end{center}
\end{adjustbox}
\end{table*}

\subsubsection{Scenario-aware model development}
% Summarize the scenario-aware pipeline for battery degradation mode estimation model development
Large amounts of battery data have been generated under well-controlled operating conditions in the past few years~\cite{mayemba2024aging}. With these high-quality laboratory data, various battery degradation models can be developed for a range of applications throughout a battery's life~\cite{sulzer2021challenge}. However, there is a lack of frameworks that enable the transfer of battery models to target scenarios different from those for which they were initially developed. To address this, we have proposed a scenario-aware pipeline to develop the best model using input features with predictive power generalized across different scenarios~\cite{zhang2024scenario}. With the aid of this pipeline, the hybrid physics-informed model for battery degradation mode estimation and phase detection is first developed in the source scenarios and then deployed in the dynamic cycling target scenario (see Fig.~\ref{fig6},~\ref{appendix_A}). 

% Offline training stage: Scenario-aware XGBoost model development for battery degradation phase detection in in the protocol cycling scenario A, DeepHPM model development for degradation mode estimation in the protocol cycling scenario B and 
At the offline training stage, the structure of surrogate NN $\mathcal{F}(\cdot)$ is set the same as that of the dynamics NNs $\mathcal{G}(\cdot)$, and the hyperbolic tangent function is used as the activation function in the DeepHPM. Then, the optimal structure of a hybrid physics-informed model is searched for using a Bayesian hyperparameter optimization framework in the source scenarios~\cite{akiba2019optuna}. Specifically, the DeepHPM was learned by minimizing the loss function defined in Eqn. (\ref{DeepHPM_loss}). Xavier Normal was used for weight initialization, and Adam was used as the optimizer in the training process. The XGBoost classifier was learned separately by minimizing the loss function defined in Eqn. (\ref{XGBoost_classifier_loss}). Since there is no cell with knee occurrence in cycling protocol B but 4 cells with knee occurrence in cycling protocol A, we use 6 cells in cycling protocol A for developing the XGBoost classifier, and 6 cells in cycling protocol B for developing the DeepHPM. Specifically, the ambient temperature is first used as the criterion to classify cells into low-temperature (10$^{\circ}$C) cells, medium-temperature (25$^{\circ}$C) cells, and high-temperature (40$^{\circ}$C) cells. Then the battery data in each cycling protocol scenario is split into a training set (3 cells) and a test set (3 cells). Notably, equal ratios of low-temperature, medium-temperature, and high-temperature cells are kept in the training and test set at each split. Finally, the XGBoost classifier and DeepHPM using the feature set with the best performance over 5 train-test splits are used for developing the final hybrid physics-informed model.
The optimal structures of the DeepHPM and XGBoost classifier can be found in Table~\ref{tab_8} and Table~\ref{tab_9},~\ref{appendix_B}.

% Online deployment stage: Implement the proposed fine-tuning strategy in the dynamic cycling scenario
At the online deployment stage, there are 2 cells with knee occurrence in the dynamic cycling scenario. Therefore, whether or not a knee occurred on the capacity fade curve is used as the criterion to first classify cells into knee-occurrence cells and no-knee-occurrence cells. Then the battery data is split with 4 cells in a training set and 2 cells in a test set. It is also ensured that there is always one cell with knee occurrence and one cell without knee occurrence in the test set at each train-test split. Then different amounts of labeled data in the dynamic cycling scenario are used to fine-tune the surrogate NN $\mathcal{F}(\cdot)$ in the pre-trained hybrid physics-informed model to determine the minimum amount of data needed for satisfactory model performance in the dynamic cycling scenario. Lastly, to reduce the randomness effect, the train-test split is repeated 5 times and the averaged experimental results are reported.    

\subsubsection{Model performance evaluation metrics}
% Model performance evaluation metrics for degradation mode estimation and phase detection.
To measure the performance of battery degradation mode estimation, the root mean square error (RMSE) metric is used, i.e.,
\begin{equation}
    \text{RMSE} = \sqrt{\frac{1}{N} \sum^{N}_{i=1}(u_i-\hat{u}_i)^2},
\end{equation}
where $u_i$ and $\hat{u}_i$ denote the estimated and observed degradation mode of sample $i$, and $N$ denotes the number of samples in the test set.
To measure the performance of battery degradation phase detection, we use four metrics, i.e., precision, recall, F1-score, and accuracy, defined as~\cite{grandini2020metrics}
\begin{align}
    \text{Precision} &=  \frac{\mathrm{TP}_k}{\mathrm{TP}_k + \mathrm{FP}_k}\\
    \text{Recall}    &= \frac{\mathrm{TP}_k}{\mathrm{TP}_k + \mathrm{FN}_k}\\
    \text{F1-score}  &= 2 \times \frac{\mathrm{Precision} \times \mathrm{Recall}}{\mathrm{Precision} + \mathrm{Recall}} \\
    \text{Accuracy}  &= \frac{1}{N} \sum^{N}_{i=1} \mathds{1}_{y_i=\hat{y}_i}    
\end{align}
where $\mathds{1}_{y_i=\hat{y}_i}$ is an indicator function that equals $1$ if the predicted class $\hat{y}_i$ is the same as the observed class $y_i$ of sample $i$ and $0$ otherwise. TP (true positives) are the samples that have been predicted as class $k$ by the model when they actually belong to class $k$, while FP (false positives) are the samples that have been predicted as class $k$ by the model when they actually belong to other classes. Similarly, TN (true negatives) and FN (false negatives) can be defined. 
The precision for class $k$ measures the proportion of correctly predicted samples as class $k$ out of the total number of samples predicted as class $k$.
The recall for class $k$ measures the proportion of correctly predicted samples as class $k$ out of the total number of samples that actually are in class $k$.
The F1-score for class $k$ aggregates precision and recall for class $k$ into a harmonic mean of both. The harmonic mean can be used to find a trade-off between precision and recall for class $k$.
The accuracy measures the proportion of correctly predicted samples of all three classes out of the total number of samples.
Precision, recall, and F1-score are calculated for each class $k$ in the test set, while the accuracy is calculated for all three classes over the entire test set.

\section{Results and discussion}
To comprehensively evaluate the proposed transferable physics-informed framework, this section is divided into three subsections. First, the battery degradation mode estimation and phase detection performance of the proposed model are evaluated in source scenarios at the offline training stage. Secondly, battery degradation mode estimation and phase detection performance of the pre-trained model and the proposed transfer learning strategy are evaluated in the target scenario at the online deployment stage. Lastly, a case study discusses advanced battery management system functions that can be enabled in a performance digital twin.
\subsection{Model evaluation in the source scenario}
%% Step 1: Specify where figures/tables are located and then followed by a summary
At the offline training stage in the source scenario, the battery degradation mode estimation results of the optimal hybrid physics-informed model using 5 different feature sets are summarized in Table~\ref{tab_4}. The model's battery degradation phase detection results are summarized in Table~\ref{tab_5}. 

\begin{table*}[htpb]
\caption{DeepHPM degradation mode estimation performance in the source scenario.}
\label{tab_4}
% \begin{adjustbox}{width=\textwidth,center}
\begin{center}
\begin{tabular}{|l|c|c|c|}
\hline
\diagbox[width=2.5\dimexpr \textwidth/8+2\tabcolsep\relax, height=1.2cm]{\textbf{Feature set}}{\textbf{Target variable}} & \textbf{LLI} & \textbf{LAM\_NE} & \textbf{LAM\_PE}\\
\hline
1D voltage-based 3-feature set & 0.0080 & 0.0124 & 0.0128\\
\hline
1D voltage-based 5-feature set & \textbf{0.0066} & 0.0123 & 0.0132\\
\hline
1D current-based 3-feature set & 0.0083 & 0.0119 & 0.0132\\
\hline
1D current-based 5-feature set & 0.0079 & \textbf{0.0118} & 0.0126\\
\hline
2D current-voltage 17-feature set & 0.0071 & 0.0140 & \textbf{0.0076} \\
\hline
\multicolumn{4}{l}{\shortstack[l]{Bold values denote the minimum RMSE in each column.}}
\end{tabular}
\end{center}
% \end{adjustbox}
\end{table*}

\begin{table*}[htpb]
\caption{XGBoost degradation phase detection performance in the source scenario.}
\label{tab_5}
% \begin{adjustbox}{width=\textwidth,center}
\begin{center}
\begin{tabular}{|l|c|c|c|}
\hline
\diagbox[width=1.2\dimexpr \textwidth/8+2\tabcolsep\relax, height=1.2cm]{\textbf{Metric}}{\textbf{Phase}} & \textbf{1} & \textbf{2} & \textbf{3}\\
\hline
Precision & 0.99 & 0.94 & 0.96\\
\hline
Recall & 0.96 & 0.98 & 0.83\\
\hline
F1-score & 0.97 & 0.96 & 0.89\\
\hline
Accuracy & \multicolumn{3}{|c|}{96\%}\\
\hline
\end{tabular}
\end{center}
% \end{adjustbox}
\end{table*}

%% Step 2: [Battery degradation mode estimation performance] Highlight typically major findings followed by minor findings
From the RMSE results in Table~\ref{tab_4} we conclude that the voltage-based 5-feature set (see Table~\ref{tab_2}) performs the best in LLI estimation, the current-based 5-feature set performs the best in LAM\_NE estimation, and current-voltage 17-feature set performs the best in LAM\_PE estimation. 
The time spent outside the upper and lower cut-off voltages, which correspond to overcharge or over-discharge conditions, can increase the likelihood of lithium-consuming degradation mechanisms like electrolyte decomposition, which irreversibly consumes lithium ions in the process of forming additional SEI layers as well as lithium plating. This may explain why the voltage-based 5-feature set performs the best in LLI estimation. In contrast, the current-based 5 feature set performs the best in LAM\_NE estimation while the current-voltage 17-feature set performs the best in LAM\_PE estimation. It can be rationalized that high currents drive rapid intercalation and deintercalation of lithium ions, which can induce alternating mechanical stress within the electrodes. Over time, these mechanical stresses can result in particle cracking and stress-driven LAM. Moreover, higher currents generate more heat inside the cell, and the resulting temperature increase can accelerate thermal-driven LAM, such as binder decomposition. Interestingly, current-based features are sufficient to achieve the best LAM estimation for the graphite-$\text{SiO}_\text{x}$ anode but the combination of current and voltage features is necessary to achieve the best LAM estimation for the NMC 811 cathode. 
However, current features alone do not provide information about the electrochemical potential of the cell and therefore cannot be independently utilized for the estimation of battery degradation modes. The voltage-based and current-voltage feature sets may be better choices for estimating degradation modes. 

%% Step 2: [Battery degradation phase detection performance] Highlight typically major findings followed by minor findings
The battery degradation phase detection performance, as measured by precision, recall, F1-score, and accuracy, are given in Table~\ref{tab_5}. The XGBoost classifier performs the best (closest to 1) in predicting Phase 1 and the least in predicting Phase 3. It can be rationalized that all the cells in the source scenario (i.e., cycling protocol A) have undergone Phase 1, and are now in either Phase 2 or 3. Consequently, the XGBoost classifier has been predominantly trained on data from Phase 1, with a smaller amount of data from Phase 2, and the least data from Phase 3. For the detection of Phase 1, the precision indicates that 99\% of the time the model correctly predicted samples as Phase 1 out of the total number of samples predicted to be in Phase 1 in the test set. The recall indicates that 96\% of the time the model correctly predicted samples as Phase 1 out of the total number of samples that actually are in Phase 1 in the test set. The F1-score can be interpreted as a weighted average between precision and recall for Phase 1, and the high value (97\%) indicates a good trade-off between precision and recall for Phase~1. Finally, the accuracy indicates that the model correctly classified samples in the test set to each of the three degradation phases 96\% of the time.

%% Step 3: Comment on implications, problems, expectations, recommendations, etc
Overall, these results indicate that the pre-trained hybrid physics-informed model is effective in estimating battery degradation modes using five histogram-based feature sets, as well as detecting degradation phases in the source scenario. In the next subsection, the pre-trained model will be deployed in the target scenario.

\subsection{Model evaluation in the target scenario}
%% Step 1: Specify where figures/tables are located and then followed by a summary
At the online deployment stage in the target scenario, the battery degradation mode estimation performance using five histogram-based feature sets and battery degradation phase detection performance of the pre-trained hybrid models are first evaluated on two cells in the test set, i.e., one cell with knee occurrence, and the other without knee occurrence. Then, the pre-trained models are fine-tuned using different amounts of labeled data, i.e., one cell with or without knee occurrence, or two cells with and without knee occurrence in the target scenario. Lastly, the fine-tuned models are evaluated using two cells in the test set. The battery degradation mode estimation results are summarized in Table~\ref{tab_6} and Table~\ref{tab_10},~\ref{appendix_B}, while the battery degradation phase detection results using estimated degradation modes as inputs are summarized in Table~\ref{tab_7}.
\begin{table*}[htpb]
\caption{DeepHPM degradation mode estimation performance in the target scenario.}
\label{tab_6}
\begin{adjustbox}{width=\textwidth,center}
% \begin{center}
\begin{tabular}{|l|c|c|c|c|}
\hline
\diagbox[width=2.5\dimexpr \textwidth/8+2\tabcolsep\relax, height=1.2cm]{\textbf{Feature set}}{\textbf{Model}} & \textbf{Pre-trained} & \shortstack[c]{\textbf{Fine-tuned with}\\ \textbf{1 cell without} \\ \textbf{knee occurrence}} & \shortstack[c]{\textbf{Fine-tuned with} \\ \textbf{1 cell with} \\ \textbf{knee occurrence}} & \shortstack[c]{\textbf{Fine-tuned with}\\\textbf{2 cells}}\\
\hline
1D voltage-based, 3-feature set & [0.0489,0.0433,0.0429] & [0.0500,0.0477,0.0478] & [0.0359,0.0325,0.0288] & [0.0231,0.0266,0.0312]\\
1D voltage-based, 5-feature set & [0.0479,0.0521,0.0434] & [0.0516,0.0502,0.0488] & [0.0361,0.0360,0.0278] & [\textbf{0.0203},0.0240, \textbf{0.0278}]\\
1D current-based, 3-feature set & [0.0531,0.0530,0.0419] & [0.0494,0.0482,0.0491] & [0.0422,0.0373,0.0333] & [0.0351,0.0305,0.0336]\\
1D current-based, 5-feature set & [0.0479,0.0461,0.0440] & [0.0503,0.0485,0.0484] & [0.0395,0.0320,0.0326] & [0.0244,0.0263,0.0317]\\
2D current-voltage, 17-feature set & [0.0764,0.0597,0.0590] & [0.0524,0.0510,0.0477] & [0.0373,0.0406,0.0443] & [0.0222,\textbf{0.0220},0.0281]\\
\hline
\multicolumn{5}{l}{\shortstack[l]{$[x, y, z]$ denotes RMSE values for LLI, LAM\_NE, and LAM\_PE, respectively. The bold values denote the minimum RMSE.}}
\end{tabular}
% \end{center}
\end{adjustbox}
\end{table*}

\begin{table*}[htpb]
\caption{XGBoost degradation phase detection performance in the target scenario.}
\label{tab_7}
\begin{adjustbox}{width=\textwidth,center}
% \begin{center}
\begin{tabular}{|l|c|c|c|c|}
\hline
\diagbox[width=1.2\dimexpr \textwidth/8+2\tabcolsep\relax, height=1.2cm]{\textbf{Metric}}{\textbf{Model}} & \textbf{Pre-trained} & \shortstack[c]{\textbf{Fine-tuned with}\\ \textbf{1 cell without} \\ \textbf{knee occurrence}} & \shortstack[c]{\textbf{Fine-tuned with} \\ \textbf{1 cell with} \\ \textbf{knee occurrence}} & \shortstack[c]{\textbf{Fine-tuned with}\\\textbf{2 cells}}\\
\hline
Precision & [0.63,0.71,0.00] & [0.97,0.76,0.00] & [0.96,0.83,1.00] & [0.97,0.83,1.00] \\
\hline
Recall & [1.00,0.69,0.00] & [0.96,0.98,0.00] & [0.95,0.98,0.40] & [0.96,0.98,0.39] \\
\hline
F1-score & [0.78,0.70,0.00] & [0.96,0.86,0.00] & [0.96,0.90,0.57] & [0.96,0.90,0.57] \\
\hline
Accuracy & 67.24\% & 82.11\% & 88.19\% & 88.29\% \\
\hline
\multicolumn{5}{l}{\shortstack[l]{$[x, y, z]$ denotes each classification metric value for three degradation phases, respectively.}}
\end{tabular}
% \end{center}
\end{adjustbox}
\end{table*}

%% Step 2: [Battery degradation mode estimation performance] Highlight typically major findings followed by minor findings
According to the results in Table~\ref{tab_6}, the hybrid model developed using the voltage-based 5 feature set and then fine-tuned using data from two cells (i.e., one with and the other without knee occurrence) in the target scenario performs the best in LLI and LAM\_PE estimation and the hybrid model developed using the current-voltage 17-feature set performs the best in LAM\_NE estimation. To retain the correlation between current and voltage in terms of time spent, the current-voltage 17-feature set is chosen to estimate different degradation modes in this target scenario and possibly in field applications as well.
In addition, a robustness analysis is also conducted, in which we simulate two cases. In the first case, two Gaussian noises, i.e., $\mathcal{N}(0, 0.04\sigma^2_{x})$ and $\mathcal{N}(0, 0.25\sigma^2_{x})$, are added separately to the test data of 2D histogram 17-feature set in the target scenario. $\sigma_x$ denotes the standard deviation per input feature estimated in the source scenario; In the second case, considering that the 1D voltage-based or current-based 5-feature sets have been demonstrated to provide better model performance than their corresponding 1D 3-feature ones (see Table~\ref{tab_6}), the test data of four histogram features in extreme ranges (i.e., $I_{01}V_{01}$, $I_{01}V_{34}$, $I_{34}V_{01}$, and $I_{34}V_{34}$ in Table~\ref{tab_2}) in the target scenario are set to zero. The robustness analysis results are reported in Table~\ref{tab_11},~\ref{appendix_B}. It can be concluded from Table~\ref{tab_11} that adding Gaussian noise makes model performance worse than missing four histogram features in extreme ranges. Thus, mitigating model performance degradation due to input noise should be prioritized in the field.
Lastly, the performance of the DeepHPM degradation mode estimation is compared against that of the Gaussian process (GP) regression surrogate model. The results in the source scenario are reported in Table~\ref{tab_12}, and the results in the target scenario are reported in Table~\ref{tab_13},~\ref{appendix_B}. It can be concluded that the GP regression surrogate model performs better than the DeepHPM model in estimating three degradation modes in the source scenario but the GP regression surrogate model pretrained in the source scenario performs worse than the fine-tuned DeepHPM model in the target scenario. This is because the standard GP regression model lacks the mechanisms to incorporate the degradation mode dynamics defined in Eqn. (\ref{eqn_u_dynamics}), which makes it difficult to transfer degradation knowledge learned in the source domains to the target domain.

%% Step 2: [Battery degradation phase detection performance] Highlight typically major findings followed by minor findings
By estimating the three degradation modes using the current-voltage 17-feature set, the battery's degradation phase can also be monitored online. The degradation phase detection performance, as measured by precision, recall, F1-score, and accuracy, can be seen in Table~\ref{tab_7} that the classification performance significantly improved after fine-tuning the model, i.e., accuracy improves from 67.24\% to 82.11\%, using one cell without knee and to 88.19\% using one with knee occurrence. Using both one cell with and one cell without knee occurrence improved accuracy only slightly more, to 88.29\%. Notably, the model did not successfully detect Phase 3 (i.e., precision, recall, and F1-score for Phase 3 are all 0.00) if fine-tuned using only one cell without knee occurrence but becomes capable if fined-tuned using one cell with knee occurrence. 

\begin{figure*}[!ht]
\centering
\subfloat[The predicted and observed degradation modes versus cumulative charge.]{\includegraphics[width=2.5in]{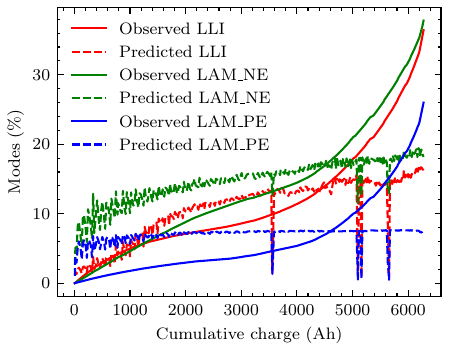}%
\label{fig7_1}}
\hfil
\subfloat[The detected and observed degradation phase versus cumulative charge.]{\includegraphics[width=2.5in]{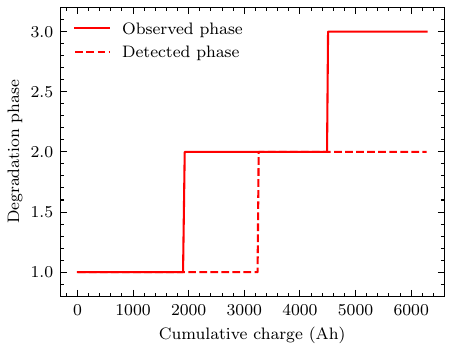}%
\label{fig7_2}}
\caption{Predicted and observed degradation modes (left) and detected and observed degradation phase (right) of a sample cell [E4C] in the test set without fine-tuning.}
\label{fig7}
\end{figure*}

\begin{figure*}[!ht]
\centering
\subfloat[The predicted and observed degradation modes versus cumulative charge.]{\includegraphics[width=2.5in]{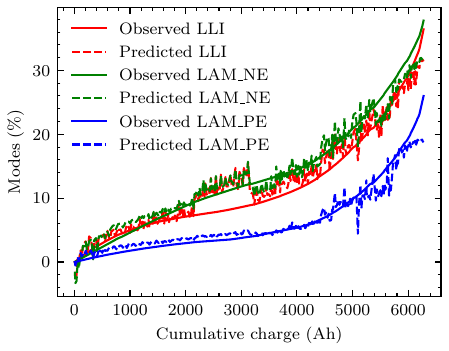}%
\label{fig8_1}}
\hfil
\subfloat[The detected and observed degradation phase versus cumulative charge.]{\includegraphics[width=2.5in]{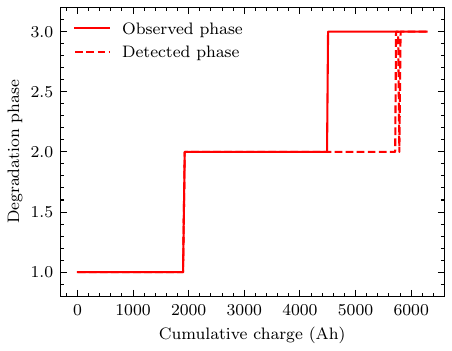}%
\label{fig8_2}}
\caption{Predicted and observed degradation modes (left) and detected and observed degradation phase (right) of a sample cell [E4C] in the test set with hybrid model fine-tuned using 1 cell with knee occurrence.}
\label{fig8}
\end{figure*}

%% Step 2: [Fine-tuning strategy effectiveness] Highlight typically major findings followed by minor findings
To demonstrate the effectiveness of the proposed fine-tuning strategy, we showcase the battery degradation mode estimation and phase detection results of a sample cell [E4C] before fine-tuning in Fig.~\ref{fig7}, and after fine-tuning using one cell with knee occurrence in Fig.~\ref{fig8}, respectively. The three degradation modes of the sample cell all begin with a square root dependence on time until they reach the knee-onset point (close to the inflection point), after which degradation modes grow exponentially. As a result, the sample cell then transfer to Phase 3. However, the pre-trained model does not predict the exponential growth of the degradation modes, and as a result, Phase 3 was not successfully detected using the estimated degradation modes (see Fig.~\ref{fig7}). After fine-tuning the pre-trained model using one cell with knee occurrence in the target scenario, the model does not only successfully predict the exponential growth of three degradation modes, but also detect Phase 3 although with some delay.
%% Step 2: Show how knee-onset points can be detected online.
With degradation phases detected with high accuracy, the knee-onset point can also be detected online as the transition point from Phase 1 to Phase 2 (see Fig.~\ref{fig8}).
%% Step 2: Show how linear correlation between knee and knee-onset can achieve early prediction of knee online
Notably, strong linear correlations were found between knee-onset and knee points identified using the curvature-based method in our previous work~\cite{zhang2024battery}. 
% As illustrated in Fig.~\ref{fig9}, 
We again find a strong linear correlation between knee-onset and knee ($\rho=0.962$) using this curvature-based identification method. With this strong linear correlation, online battery capacity knee prediction can be made from detected knee-onset points.

% \begin{figure}[!ht]
% \centerline{\includegraphics[width=0.5\textwidth]{./Images/ICL_NMC_knee_kneeonset_Ah.pdf}}
% \caption{The relationship between knee-onset and knee for 6 cells with knee occurrence.}
% \label{fig9}
% \end{figure}

%% Step 3: Comment on implications, problems, expectations, recommendations, etc
Lastly, based on the results reported in Table~\ref{tab_6} and Table~\ref{tab_7}, it can be concluded that the effectiveness of the proposed fine-tuning strategy highly depends on whether or not the fine-tuning cell has a capacity knee or not in the target scenario. Moreover, using more cells in the target scenario to fine-tune the model does improve the model performance further. However, considering that labeled data is often scarce in field applications, it can be essential to use a minimum amount of labeled data required to achieve sufficiently high model accuracy. Using data of only one cell with knee occurrence would set a lower bar for this.

In Ref.~\cite{wang2024physics}, a physics-informed neural network (PINN) is proposed for battery state-of-health (SoH) estimation, and a fine-tuning strategy is proposed to transfer the PINN model for SoH estimation of batteries with different chemistries and charge/discharge protocols. The key differences between Ref.~\cite{wang2024physics} and our work are specified as follows: 1) Problems \& Models: The PINN is used to capture battery capacity fade dynamics and estimate battery state-of-health (SoH) in Ref.~\cite{wang2024physics}, while in our work, the PINN is used to capture battery degradation mode dynamics, i.e., loss of lithium inventory (LLI) and loss of active material (LAM), and its estimated degradation modes are then taken as inputs to an XGBoost model to perform battery degradation phase detection and online battery classification; 2) Input features \& Data: The cycle-based features, i.e., mean, standard deviation, kurtosis, skewness, charging time, accumulated charge, curve slope, and curve entropy, extracted from charge data before the battery is fully charged, are used as the inputs to the PINN for SoH estimation in Ref.~\cite{wang2024physics}, while the histogram-based features, i.e., time spent within specific voltage and/or current ranges, extracted from full charge-discharge data regardless of battery usage, are used as the inputs to the PINN for degradation mode estimation in our work; 3) Outputs: The output of the PINN model is battery capacity in Ref.~\cite{wang2024physics}, while the outputs of our model (PINN+XGBoost) are three degradation modes, degradation phase. With accurately detected degradation phase, the knee-onset point can also be detected online as the transition point from Phase 1 to Phase 2. 
The advantages of our method are as follows: 1) The capacity fade is attributed to the growth of degradation modes (LLI and LAM). Our method is capable of estimating these degradation modes, which provides diagnostic information for the occurrence of battery capacity knee. With these estimated degradation modes, degradation phases and knee-onset point can be detected in the target scenario (CC-CV charge/WLTP driving discharge profiles) with high accuracy; 2) Instead of extracting from either charge or discharge data, the 2D histogram features are extracted from full charge-discharge data regardless of battery usage, and are proven to be effective in estimating battery degradation modes, detecting degradation phase and knee-onset point.

Transformer-based models have shown excellent performance of estimating battery SoH in recent studies~\cite{xu2023hybrid}\cite{zhao2025predictive}. However, the proposed hybrid physics-informed model and Transformer-based models offer distinct advantages in battery SoH estimation and degradation diagnosis. The hybrid physics-informed model aims to embed battery degradation dynamics into the neural network (NN) architecture, leveraging the advantages of both, i.e., the flexibility of NNs and the interpretability of physics. In contrast, Transformer-based models excel in battery SoH estimation accuracy without physical insights. The choice between the hybrid physics-informed and Transformer-based models depends on the specific requirements in battery applications.

\subsection{Computational requirements}
The offline training and online deployment of the hybrid model (DeepHPM and XGBoost) are performed on a desktop (OS: Windows 10, Memory: 64 GB, CPU: Intel Core i7-13700, GPU: NVIDIA GeForce RTX 3070). 
At the offline training stage, the search time for the optimal hybrid model structure using one histogram-based feature set is approximately 10 hours. The optimal structure of the hybrid model can be found in Table~\ref{tab_8} and Table~\ref{tab_9},~\ref{appendix_B}. The training time for the hybrid model using one histogram-based feature set is approximately 4 minutes.
At the online deployment stage, the fine-tuning time to create one local model in the target scenario is approximately 30 seconds. The test (or inference) time per sample is in milliseconds, which is negligible compared to the battery degradation rate, the sampling rate, and the histogram aggregation rate.

\subsection{A case study: advanced battery management system functions in a performance digital twin}
% Choose one of the following application cases~\cite{aykol2021perspective}:
% \begin{itemize}
% \item The prediction performance can potentially be further improved.
% \item We expect improvements in model performance under unseen parameters, which consequently reduce the overall number of parameters tested with a factor of 5.
% \item The number of experimental cycles required to predict lifetime can be reduced to only the first 10 cycles.
% \item The number of experimental cycles required to classify charging policies or sort cells can be reduced to less than 5 cycles.
% \item  The benefits of improved on-board lifetime prediction and other on-vehicle applications.
% \end{itemize}
The histogram-based feature engineering method, hybrid physics-informed model, and the proposed fine-tuning strategy are key enablers for a concept of battery performance digital twin (PDT), or cloud battery management system (BMS)~\cite{naseri2023digital}. They can enable advanced BMS functionalities, such as online degradation diagnosis and prognosis, aging-aware battery classification, and second-life repurposing. 
% as illustrated in Fig.~\ref{fig10}. 
Time-series voltage and current data are commonly measured for all cells connected in series inside a battery pack. The onboard BMS can first aggregate these time-series voltage and current data as histograms and then communicate to the PDT on request or at a very slow sampling rate via the Internet-of-Things (IoT) gateway~\cite{naseri2023digital}. The global battery PDT, or the hybrid physics-informed model in this work, first uses a small amount of labeled data to create a local PDT in a target scenario. The local PDT is then used to estimate degradation modes and detect the degradation phase for each cell in a battery pack. Based on cell-level estimated degradation modes and degradation phase, the cell-to-cell heterogeneity inside a battery pack can be determined for aging-aware battery classification later on. For example, the sample cell [E4C] in Fig.~\ref{fig8} was detected to be in Phase 3. The pack or module where the cell is located may be either repurposed to second-life applications in which the knee occurrence can be stopped, or be recycled. The specific second-life repurposing, however, also requires additional information, such as battery energy and power capabilities, technical requirements of second-life applications, residual value estimation, etc~\cite{mathews2020technoeconomic}.

% \begin{figure}[!ht]
% \centerline{\includegraphics[width=\textwidth]{./Images/The_physical_battery_and_its_digital_twin_in_the_cloud.pdf}}
% \caption{The physical battery and its performance digital twin in the cloud.}
% \label{fig10}
% \end{figure}

\section{Conclusions}
% Step 1 [optional] - Background information (purpose of this work, proposed methodology)
To alleviate the technical, economic, and safety concerns arising from capacity knee occurrence during the service life of a battery, a transferable physics-informed framework that consists of a histogram-based feature engineering method, a hybrid physics-informed model, and a fine-tuning strategy was proposed for online battery degradation diagnosis and knee-onset detection. 
% Step 2 [compulsary] - Summarizing and commenting on the key results (making claims, explaining the results, comparing this work with previous ones, indicating application values, etc.)
Specifically, pre-trained hybrid physics-informed models were first developed using 1D or 2D histogram-based feature sets and their battery degradation mode estimation and phase detection performance were evaluated in the source scenarios using a scenario-aware pipeline. The pre-trained hybrid models were then fine-tuned using different amounts of labeled data, and deployed in the target scenario. Among the five histogram-based feature sets investigated, it was demonstrated that the 2D histogram-based 17-feature set was the best for battery degradation mode estimation in both source and target scenarios, here and possibly in field applications as well. The fine-tuning strategy was proven to be effective in improving not only battery degradation mode estimation but also degradation phase detection using 3 estimated degradation modes in the target scenario. With degradation phases detected with high accuracy, online prediction of battery capacity knee points can also be achieved by leveraging the strong linear correlation identified between knee-onset and knee points. Lastly, it has been found that using one cell with knee occurrence in the target scenario may be enough to achieve a satisfactory model accuracy. 

% Step3 [compulsary] - Stating the limitations of the current work and making recommendations for future work (avoid saying what we will do)
As key enablers for the concept of battery performance digital twin (DPT) in the cloud, the proposed framework can enable advanced BMS functions, such as online degradation diagnosis and prognosis, aging-aware battery classification, and second-life repurposing. As a result, the overall value of electric vehicle batteries can be maximized before recycling. In terms of future work, 1) it would be interesting to quantify the aleatoric uncertainty arising from the noisy data and the epistemic uncertainty from the model structure using the Bayesian approach; 2) we will further evaluate the performance of the proposed framework in real-world applications (e.g., electric vehicles and grid storage) once battery data in the field are available; 3) Considering abundant unlabeled data in the field that poses new challenges to the fine-tuning strategy. Active learning algorithms that iteratively expand the labeled data in the lab will be investigated; 4) For different types of batteries or accelerated aging mechanisms, we will conduct a systematic study to quantify the accuracy loss of the proposed fine-tuning strategy once degradation modes of other types of batteries or accelerated aging mechanisms become available.

\section*{CRediT authorship contribution statement}
\textbf{Huang Zhang:} Conceptualization, Methodology, Software, Validation, Formal analysis, Data curation, Writing – original draft, Project administration, Funding acquisition. \textbf{Xixi Liu:} Resources, Writing – review \& editing. \textbf{Faisal Altaf:} Resources, Writing – review \& editing, Supervision, Funding acquisition. \textbf{Torsten Wik:} Resources, Writing – review \& editing, Supervision, Funding acquisition.

\section*{Declaration of competing interest}
 The authors declare that they have no known competing financial interests or personal relationships that could have appeared to influence the work reported in this paper.

\section*{Acknowledgements}
This work was supported by the Swedish Energy Agency (Grant number P2024-00998).

\appendix{}
\section{Supplementary figures}\label{appendix_A}
\begin{figure}[!ht]
\centerline{\includegraphics[width=0.75\textwidth]{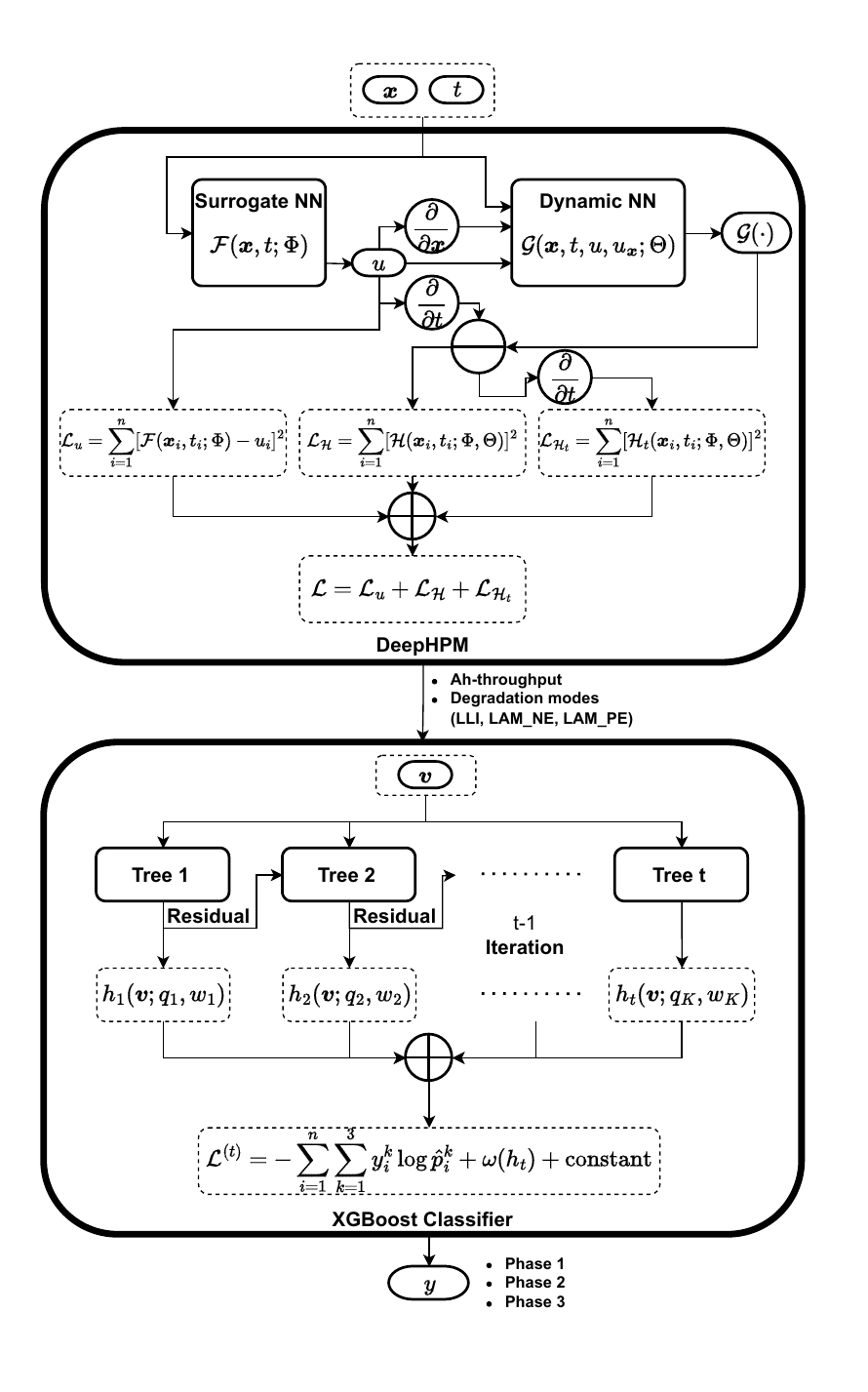}}
\caption{Hybrid physics-informed model.}
\label{fig4}
\end{figure}

\begin{figure*}[!ht]
\centerline{\includegraphics[width=1.3\textwidth]{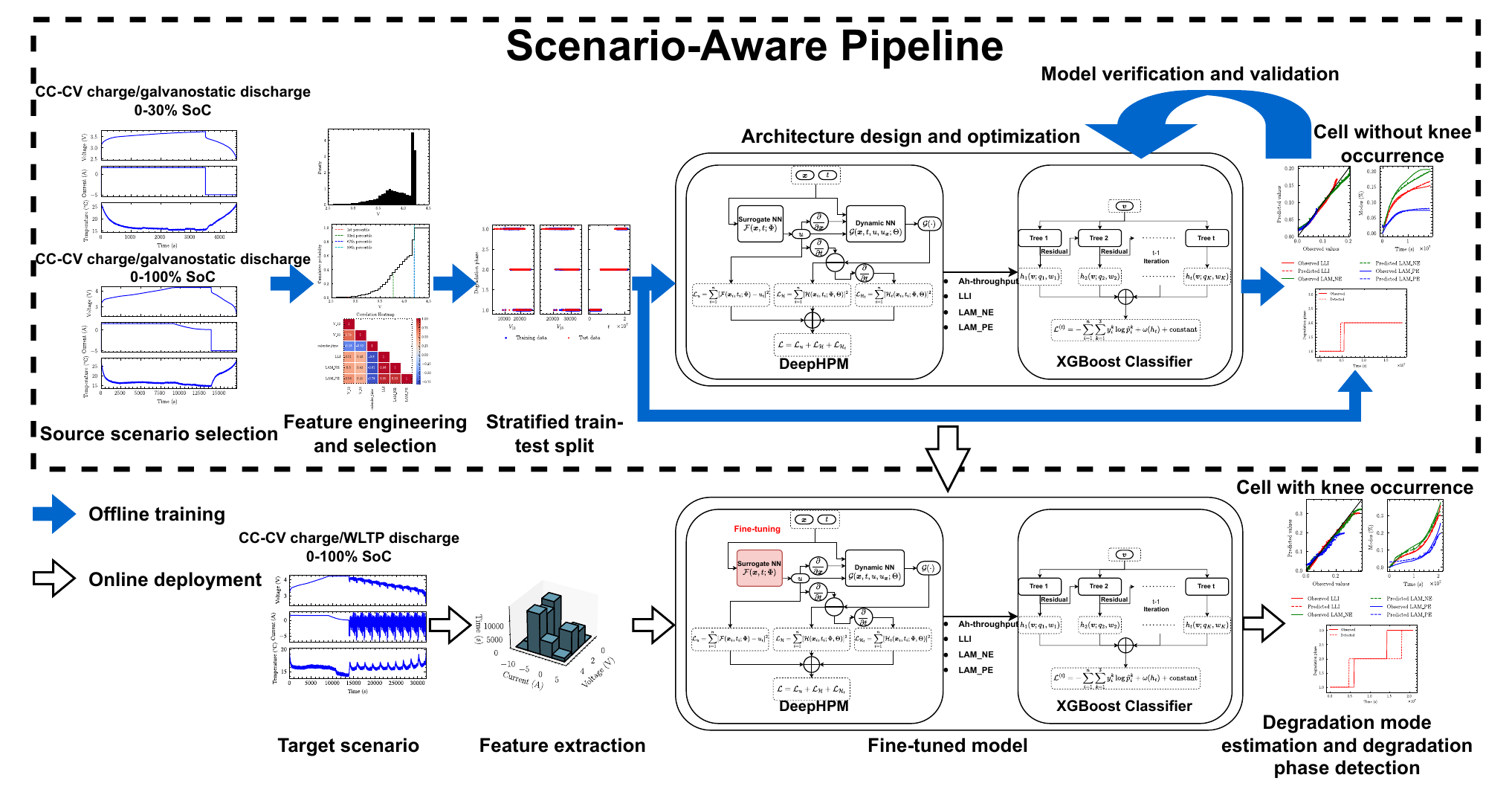}}
\caption{The scenario-aware model development pipeline for degradation mode estimation and phase detection.}
\label{fig6}
\end{figure*}

\FloatBarrier
\section{Supplementary tables}\label{appendix_B}
\begin{table*}[htpb]
\caption{Optimal DeepHPM structures for battery degradation mode estimation in the source scenario.}
\label{tab_8}
% \begin{adjustbox}{width=\textwidth,center}
\begin{center}
\begin{tabular}{|l|c|c|c|}
\hline
\diagbox[width=2.5\dimexpr \textwidth/8+2\tabcolsep\relax, height=1.2cm]{\textbf{Feature set}}{\textbf{Target variable}} & \textbf{LLI} & \textbf{LAM\_NE} & \textbf{LAM\_PE}\\
\hline
1D voltage-based 3-feature set & [2, 64] & [6, 32] & [4, 64]\\
\hline
1D voltage-based 5-feature set & [2, 56] & [2, 56] & [6, 48]\\
\hline
1D current-based 3-feature set & [4, 64] & [2, 48] & [8, 40]\\
\hline
1D current-based 5-feature set & [4, 56] & [2, 40] & [6, 64]\\
\hline
2D current-voltage 17-feature set & [2, 32] & [2, 32] & [4, 48] \\
\hline
\multicolumn{4}{l}{\shortstack[l]{$[x, y]$ denotes the number of hidden layers and neurons per layer.}}
\end{tabular}
\end{center}
% \end{adjustbox}
\end{table*}

\begin{table*}[ht]
\caption{Optimal XGBoost structures for battery degradation phase detection in the source scenario.}
\label{tab_9}
% \begin{adjustbox}{width=\textwidth,center}
\begin{center}
\begin{tabular}{|l|c|}
\hline
\textbf{Hyperparameter} & \textbf{Value}\\
\hline
Number of trees & 100 \\
\hline
Learning rate   & 0.0207 \\
\hline
Maximum depth of a tree & 8 \\
\hline
Minimum sum of instance weight & 7 \\
\hline
\end{tabular}
\end{center}
% \end{adjustbox}
\end{table*}

\begin{table*}[htpb]
\caption{DeepHPM degradation mode estimation performance in the target scenario.}
\label{tab_10}
\begin{adjustbox}{width=\textwidth,center}
% \begin{center}
\begin{tabular}{|l|c|c|c|c|}
\hline
\diagbox[width=2.5\dimexpr \textwidth/8+2\tabcolsep\relax, height=1.2cm]{\textbf{Feature set}}{\textbf{Model}} & \textbf{Pre-trained} & \shortstack[c]{\textbf{Fine-tuned with}\\ \textbf{1 cell without} \\ \textbf{knee occurrence}} & \shortstack[c]{\textbf{Fine-tuned with} \\ \textbf{1 cell with} \\ \textbf{knee occurrence}} & \shortstack[c]{\textbf{Fine-tuned with}\\\textbf{2 cells}}\\
\hline
1D voltage-based, 3-feature set & [0.0465,0.0413,0.0423] & [0.0457,0.0457,0.0423] & [0.0329,0.0324,0.0269] & [0.0228,0.0247,0.0307]\\
1D voltage-based, 5-feature set & [0.0474,0.0454,0.0413] & [0.0465,0.0470,0.0426] & [0.0357,0.0343,0.0269] & [0.0218,\textbf{0.0198},\textbf{0.0261}]\\
1D current-based, 3-feature set & [0.0468,0.0377,0.0401] & [0.0456,0.0451,0.0428] & [0.0386,0.0359,0.0304] & [0.0337,0.0327,0.0304]\\
1D current-based, 5-feature set & [0.0442,0.0385,0.0425] & [0.0459,0.0447,0.0428] & [0.0362,0.0302,0.0300] & [0.0256,0.0268,0.0276]\\
2D current-voltage, 17-feature set & [0.0670,0.0595,0.0570] & [0.0478,0.0482,0.0426] & [0.0381,0.0404,0.0371] & [\textbf{0.0214},0.0210,0.0265]\\
\hline
\multicolumn{5}{l}{\shortstack[l]{$[x, y, z]$ denotes standard deviation values for LLI, LAM\_NE, and LAM\_PE, respectively. The bold values denote the minimum standard deviation.}}
\end{tabular}
% \end{center}
\end{adjustbox}
\end{table*}

\begin{table*}[htpb]
\caption{Robustness analysis of DeepHPM degradation mode estimation in the target scenario.}
\label{tab_11}
\begin{adjustbox}{width=\textwidth,center}
% \begin{center}
\begin{tabular}{|l|c|c|c|c|}
\hline
\diagbox[width=2.5\dimexpr \textwidth/8+2\tabcolsep\relax, height=1.2cm]{\textbf{Case}}{\textbf{Model}} & \textbf{Pre-trained} & \shortstack[c]{\textbf{Fine-tuned with}\\ \textbf{1 cell without} \\ \textbf{knee occurrence}} & \shortstack[c]{\textbf{Fine-tuned with} \\ \textbf{1 cell with} \\ \textbf{knee occurrence}} & \shortstack[c]{\textbf{Fine-tuned with}\\\textbf{2 cells}}\\
\hline
2D current-voltage, 17-feature set & [0.0764,0.0597,0.0590] & [0.0524,0.0510,0.0477] & [0.0373,0.0406,0.0443] & [\textbf{0.0222},\textbf{0.0220},\textbf{0.0281}]\\
Additive Gaussian noise, $\mathcal{N}(0, 0.04\sigma^2_{x})$ & [0.1061,0.0723,0.0672] & [0.0534,0.0532,0.0476] & [0.0538,0.0630,0.0481] & [0.0269,0.0267,0.0297]\\
Additive Gaussian noise, $\mathcal{N}(0, 0.25\sigma^2_{x})$ & [0.1101,0.0791,0.0682] & [0.0568,0.0592,0.0487] & [0.0677,0.0772,0.0598] & [0.0395,0.0407,0.0358]\\
Missing histogram data & [0.0768,0.0589,0.0583] & [0.0518,0.0531,0.0475] & [0.0473,0.0299,0.0570] & [0.0237,0.0245,0.0282]\\
\hline
\multicolumn{5}{l}{\shortstack[l]{$[x, y, z]$ denotes RMSE values for LLI, LAM\_NE, and LAM\_PE, respectively. The bold values denote the minimum RMSE.\\ $\sigma_x$ denotes the standard deviation per input feature estimated in the source scenario.}}
\end{tabular}
% \end{center}
\end{adjustbox}
\end{table*}

\begin{table*}[htpb]
\caption{Gaussian process regression degradation mode estimation performance in the source scenario.}
\label{tab_12}
% \begin{adjustbox}{width=\textwidth,center}
\begin{center}
\begin{tabular}{|l|c|c|c|}
\hline
\diagbox[width=2.5\dimexpr \textwidth/8+2\tabcolsep\relax, height=1.2cm]{\textbf{Feature set}}{\textbf{Target variable}} & \textbf{LLI} & \textbf{LAM\_NE} & \textbf{LAM\_PE}\\
\hline
1D voltage-based 3-feature set    & 0.0061 & 0.0119 & \textbf{0.0067} \\
\hline
1D voltage-based 5-feature set    & \textbf{0.0046} & 0.0135 & 0.0075\\
\hline
1D current-based 3-feature set    & 0.0069 & \textbf{0.0093} & 0.0084\\
\hline
1D current-based 5-feature set    & 0.0078 & 0.0094 & 0.0086\\
\hline
2D current-voltage 17-feature set & \textbf{0.0046} & 0.0138 & 0.0073 \\
\hline
\multicolumn{4}{l}{\shortstack[l]{Bold values denote the minimum RMSE in each column.}}
\end{tabular}
\end{center}
% \end{adjustbox}
\end{table*}

\begin{table*}[htpb]
\caption{Gaussian process regression degradation mode estimation performance in the target scenario.}
\label{tab_13}
% \begin{adjustbox}{width=\textwidth,center}
\begin{center}
\begin{tabular}{|l|c|}
\hline
\diagbox[width=2.5\dimexpr \textwidth/8+2\tabcolsep\relax, height=1.2cm]{\textbf{Feature set}}{\textbf{Model}} & \textbf{Pre-trained}\\
\hline
1D voltage-based, 3-feature set & [\textbf{0.0359},\textbf{0.0328},0.0347] \\
1D voltage-based, 5-feature set & [0.0395,0.0380,\textbf{0.0337}] \\
1D current-based, 3-feature set & [0.0407,0.0422,0.0441] \\
1D current-based, 5-feature set & [0.0501,0.0525,0.0481] \\
2D current-voltage, 17-feature set & [0.0368,0.0333,0.0377]\\
\hline
\multicolumn{2}{l}{\shortstack[l]{$[x, y, z]$ denotes RMSE values for LLI, LAM\_NE, and LAM\_PE, respectively. \\The bold values denote the minimum RMSE.}}
\end{tabular}
\end{center}
% \end{adjustbox}
\end{table*}

\FloatBarrier
\bibliographystyle{elsarticle-num} 
\bibliography{References}





% \end{thebibliography}
\end{document}